# Journal of Geophysical Research: Space Physics

# Characterizing the ionospheric current pattern response to southward and northward IMF turnings with dynamical SuperMAG correlation networks

J. Dods[1], S. C. Chapman[1], and J. W. Gjerloev[2,3]

[1]Centre for Fusion, Space and Astrophysics, Department of Physics, University of Warwick, Coventry, UK, [2]Department of Physics and Technology, University of Bergen, Bergen, Norway, [3]The Johns Hopkins University Applied Physics Laboratory, Laurel, Maryland, USA

**Abstract** We characterize the response of the quiet time (no substorms or storms) large-scale ionospheric transient equivalent currents to north-south and south-north IMF turnings by using a dynamical network of ground-based magnetometers. Canonical correlation between all pairs of SuperMAG magnetometer stations in the Northern Hemisphere (magnetic latitude (MLAT) 50–82°) is used to establish the extent of near-simultaneous magnetic response between regions of magnetic local time-MLAT. Parameters and maps that describe spatial-temporal correlation are used to characterize the system and its response to the turnings aggregated over several hundred events. We find that regions that experience large increases in correlation post turning coincide with typical locations of a two-cell convection system and are influenced by the interplanetary magnetic field $B_y$. The time between the turnings reaching the magnetopause and a network response is found to be ~8–10 min and correlation in the dayside occurs 2–8 min before that in the nightside.

## 1. Introduction

During periods of southward directed IMF a two-cell ionospheric convection system is typically established [*Dungey*, 1961; *Lu et al.*, 2002a]. The convection system has a more complex structure during extended periods of northward IMF but normally involves a distorted two-cell or a multicell configuration [*Greenwald et al.*, 1995]. A topical question is the nature of the dynamic ionospheric convection response to an IMF turning that leads to a transition between these two states. The state of the convection system and its transition can be estimated by a variety of methods based on inferring ground magnetometer measurements or directly measuring ion flows [*Brekke et al.*, 1973; *Feldstein and Levitin*, 1986]. These methods require assumptions including the conductivity distribution in the ionosphere. The assimilative mapping of ionospheric electrodynamics (AMIE) method combines a variety of direct and indirect observations to constrain some, but not all, of these assumptions [*Richmond*, 1992].

How fast the response to an IMF turning propagates from the dayside to the nightside of the ionosphere has been the topic of considerable discussion [*Lockwood and Cowley*, 1999]. Some studies suggest a propagation time in the convective response of ~2 min/h of magnetic local time (MLT) [*Lockwood et al.*, 1986; *Todd et al.*, 1988; *Khan and Cowley*, 1999; *Fiori et al.*, 2012]. They proposed that the change occurs slowly due to the fly wheel effect. Large convective bulk flows of charged particles occur during southward periods, the charged particles collide with neutrals in the atmosphere causing neutral flows in the same direction over time. When the IMF turns northward, these $E \times B$ flows should stop but the momentum of the neutral flows prevents this from occurring immediately. Others indicate a near-simultaneous (<2 min) global response to the turnings [*Ridley et al.*, 1997, 1998; *Ruohoniemi and Greenwald*, 1998; *Yu and Ridley*, 2009]. *Lu et al.* [2002b] suggest that the situation may be a combination of the two; i.e., there is an initial magnetoacoustic wave launched as a result of the turning that rapidly initiates the convective response globally. This is followed by a slower evolution of the convection system that reaches its peak in the dayside before the nightside, accounting for the delay in convection seen in *Lockwood et al.* [1986], *Todd et al.* [1988], and *Khan and Cowley* [1999].

In the last two decades there has been a steady improvement in the global coverage of spatially distributed observations, with the advent of SuperDARN [*Chisham et al.*, 2007], SuperMAG [*Gjerloev*, 2009], and AMPERE [*Anderson et al.*, 2002; *Clausen et al.*, 2012]. However, all such ground and space based measurements are affected by the following: first, the motion of ground-based stations, due to the rotation of the Earth, inherently convolves the observed spatial and temporal timescales. Second, the spatial coverage of the measurement is nonuniform, and large regions are left without coverage; thus, for a given event the key physics may simply be missed. Finally, the complex response of the Sun-magnetosphere-ionosphere system to the varying solar wind driver depends in part on the current state, and hence, past history, of the magnetosphere itself. Hence, our results which aggregate sets of observations selected on the basis of solar wind conditions (here IMF $B_z$ turnings) will also aggregate the effect upon the response arising from different past magnetospheric time histories. The aggregation of data over many similar events to analyze the responses of the ionospheric current system to external forcing has been explored previously [*Ridley et al.*, 1998]. However, the standard approach is to essentially interpolate the sparsely distributed additional observations onto a grid, with the additional possibility of combining with model outputs [e.g., *Waters et al.*, 2015], such that there are no missing grid data point at any given time instance. An alternative approach is to characterize the spatiotemporal correlation pattern for each event directly from the spatially nonuniform original observations, then aggregate many such patterns onto a single grid to give a complete spatial coverage. The advantage is that the spatial characteristics are obtained at the highest spatial resolution. This is effectively the approach used in the pioneering work of *Heppner and Maynard* [1987] and *Iijima and Potemra* [1976]. This is the approach we take in this paper.

Here we will utilize the full set of ground-based magnetometers in the northern Hemisphere to characterize the time-dependent correlated magnetospheric/ionospheric response which in turn reflects the aspect of the convection system. We will develop a method — network analysis — that does not require assumptions about the state of the ionosphere such as conductivity but utilizes the spatiotemporal correlation between all available pairs of magnetometer stations to characterize the system response. We construct a time-dependent cross-correlation matrix at zero lag, hereafter correlation, by performing a windowed canonical correlation between the full set of SuperMAG magnetometer stations vector time series [see *Dods et al.*, 2015]. This correlation matrix can then be represented by a dynamical network. The network is then mapped onto a regular grid. We aggregate the cross-correlation network responses over 300–400 similar events to obtain an averaged response as a function of magnetic local time-magnetic latitude (MLT-MLAT) and of the time delay since the occurrence of the IMF north-south and south-north turnings. Our aggregate data set is restricted to turnings that occur during quiet time conditions, that is, |DST| < 30 nT and when there are no identifiable substorms. We propose network parameters that capture the spatial distribution of regions of enhanced cross correlation in MLT-MLAT and use them to characterize the transient equivalent currents.

The paper is organized as follows: the data sets used are detailed in section 2. In section 3 the methodology is summarized, with details given in Appendix A. In section 4 the main results are presented, additional results which group turnings for summer and winter centered epochs and those grouped based on the strength of the turnings are included in sections 4.3 and 4.4, respectively.

## 2. Data Used in This Study

Vector magnetometer time series at 1 min cadence from years 1998 to 2004 from the SuperMAG database of ground station magnetometers are used. The magnetometer time series have been preprocessed as in *Gjerloev* [2012]. All magnetometer stations between 50° and 82° MLAT were included. Station pairs that are within a 300 km geodesic separation of each other were excluded as they are likely to be correlated with each other at all times; i.e., they do not constitute spatially distinct observations. We use solar wind data from the ACE spacecraft (1998–2004) at 1 min cadence from the SuperMAG database. The data is in a preprocessed format having been propagated forward to the front of the magnetosphere using the pseudo-minimum variance technique of *Weimer et al.* [2003] and *Weimer* [2004]. We aim to characterize transitions in the quiet time transient equivalent currents using just the ground station magnetometers. Quiet time here refers to times in which no geomagnetic substorms or storms (|DST| < 30 nT) are occurring. We used the onset times from the substorm event list provided by SuperMAG to identify substorm intervals [see *Newell and Gjerloev*, 2011a, 2011b]. Data from 30 min before each substorm onset to 3 h after each substorm onset were excluded from the analysis.

North-south and south-north IMF turnings are identified as periods in which IMF $B_z$ is continually positive or negative for at least 30 min preceding the turning. We select events for which $B_z$ is continuously negative (or positive) for at least 40 min and no longer than 80 min post turning. In addition, we provide results for other intervals of continuously negative or positive $B_z$ post turning: $10-20$, $20-40$, and $80-150$ min. Continually negative (or positive) means here that $B_z$ is of the same sign for 90% of the minimum of the time interval, we exclude events that do not satisfy this criterion. We further divide the north-south and south-north sets of turning events into those for which $B_y$ is also continually positive (or negative) during the same post turning interval. We do not have any requirement on IMF $B_y$ before the north-south and south-north turning occurs. These sets of events are then aggregated; that is, the network connection values at each grid point at each time with respect to the IMF turning time are first calculated for each event then averaged over all events to provide a time-dependent map of the network.

## 3. Forming Dynamical Networks on a Fixed MLT-MLAT Grid

For studies of individual events such as substorms we can use the stations as locations of the network nodes directly [*Dods et al.*, 2015] and in this case the network nodes rotate with the Earth and are not stationary in magnetic local time (MLT) (12 o'clock MLT is defined by the Earth-Sun line). Here we will determine the average response by aggregating over many IMF turnings. The first step, for each event, is to interpolate the correlation properties of nodes in the station network, $T_{ij}$, onto a common grid that is static in MLT-MLAT. The outline of the procedure is as follows:

1. Canonical correlation [*Brillinger*, 1975] (with a 48 min running window) is used to quantify the correlation between the magnetic vector time series between each pair of stations $i$ and $j$. Canonical correlation applies a rotation to the data set such that the correlation is maximized in the first canonical component. The correlation is calculated at 2 min intervals for the years 1998–2004 at zero lag. We need to optimize for the shortest possible correlation window to give the highest time resolution and optimize for the largest number of data points within the window to minimize noise in order to calculate correlation. A weighted adjacency matrix, $C_{ij}$, is formed which contains the first canonical coefficients which quantify the correlation between the station pairs. Station-specific thresholds are applied to $C_{ij}$ to determine whether a given station pair is connected or not. If the correlation between the two stations exceeds a threshold, then the stations are connected. As in *Dods et al.* [2015], station specific thresholds are calculated from the data to normalize for the specifics of a given station that may affect its average correlation with all other stations; that is, the local ground conductivity, differing magnetometer responses, and ranges. Following normalization, all stations have the same average likelihood of being connected to the network. In *Dods et al.* [2015] we chose a normalized degree, that is, the number of connections a particular station has to the rest of the network normalized by the number of possible connections, $n_0 = 0.05$, to minimize the number of noise-related connections. Here we use $n_0 = 0.075$, so that more noise-related connections will be present in each network, but averaging the networks of many events will average out the influence of these noise-related connections. Applying the threshold then converts this correlation matrix into an intermediate network $T_{ij}$ which is a binary matrix where $T_{ij} = 1$ indicates that station $i$ is connected to station $j$ and $T_{ij} = 0$ otherwise ($T_{ij}$ is equivalent to the network formed in *Dods et al.* [2015] but for $n_0 = 0.075$).

2. We define a regular grid in MLT and MLAT extending from 50° to 82° MLAT with each grid cell spanning $\Delta\phi = 2$ h in MLT and $\Delta\theta = 8°$ in MLAT. It is usually optimal for the grids representing nodes in spatially embedded networks to have equal physical area [*Heitzig et al.*, 2012]. However, such a grid is not appropriate in this case since the lower latitude stations spatially sample a larger physical area than those at higher latitudes within the time window. Missing connections occur when pairs of MLT-MLAT grid cells are never populated by observing stations. We chose an optimal grid that gives the highest spatial resolution without missing connections.

3. Network connections are mapped onto the grid using a modified form of the Cloud-in-a-Cell scheme [*Hockney and Eastwood*, 1988]. The mapping generates matrices $A_{kl}^*(t)$ and $A_{kl}^{**}(t)$: $A_{kl}^*$ is a normalization matrix and is the connectivity of grid cell pairs $kl$ that would be obtained when all station pairs in the intermediate network are connected. $A_{kl}^{**}$ is the connectivity of grid cells $kl$ given by the time-dependent $T_{ij}$ network. Both $A_{kl}^*$ and $A_{kl}^{**}$ can take values $\geq 0$. The mapping scheme is described in detail in Appendix A.

### 3.1. Aggregate Networks and Parameters

We used the above method to aggregate $M$ north-south (and south-north) turning events to form a normalized network response matrix $A_{kl}(\tau)$. The number of events in the 40–80 min interval range north-south turning $+B_y$ set is 384, in the $-B_y$ set is 364, and in the south-north turning $+B_y$ and $-B_y$ sets are 360 and 357, respectively. We defined the event-specific network response matrices $A^*_{kl(q)}(\tau)$ and $A^{**}_{kl(q)}(\tau)$ which are time centered on the turning event $q$. The $\tau$ is the time that has elapsed following the turning. The normalized aggregate network response matrix $A_{kl}$ is given by

$$A_{kl}(\tau) = \frac{\sum\limits_{q=1}^{M} A^*_{kl(q)}(\tau)}{\sum\limits_{q=1}^{M} A^{**}_{kl(q)}(\tau)}. \tag{1}$$

The $\sum\limits_{q=1}^{M} A^{**}_{kl(q)}(\tau)$ is the maximum possible connectivity that can occur, summed over all turning events $M$; it acts as a normalization since $\sum\limits_{q=1}^{M} A^*_{kl(q)}(\tau) \leq \sum\limits_{q=1}^{M} A^{**}_{kl(q)}(\tau)$. $A_{kl}$ can then take values between 0 and 1, if $A_{kl} = 0$ then there were no network connections between grid cell pair $k, l$ for any of the events, and if $A_{kl} = 1$ there is always a connection between grid cell pair $k, l$ in every event.

We used time-varying network parameters to quantify the aggregated network spatial pattern of correlation; these are those defined in *Dods et al.* [2015] except that they differ in their normalization. Here we normalize to the maximum possible connectivity in our network $A^{**}_{kl}$ to obtain parameters as follows.

The average global connection likelihood $\alpha$ is

$$\alpha(\tau) = \frac{\sum\limits_{k=1, k\neq l}^{N_g} \sum\limits_{l=1, l\neq k}^{N_g} \sum\limits_{q=1}^{M} A^*_{kl(q)}(\tau)}{\sum\limits_{k=1, k\neq l}^{N_g} \sum\limits_{l=1, l\neq k}^{N_g} \sum\limits_{q=1}^{M} A^{**}_{kl(q)}(\tau)}, \tag{2}$$

where $N_g$ is the number of grid cells in the network. We can define parameters for connectivity between different regions of MLT and MLAT by choosing to sum over a subset of $k$ and $l$ in equation (2).

The dayside ($\phi_d$) (6 < MLT < 18), nightside ($\phi_n$) (MLT > 18 or MLT < 6) and cross-connection likelihood ($\phi_c$) are

$$\phi_{d,n,c}(\tau) = \frac{\sum\limits_{k=1, k\neq l}^{N_g} \sum\limits_{l=1, l\neq k}^{N_g} \sum\limits_{q=1}^{M} A^*_{kl(q)}(\tau) S_{kl,d,n,c}}{\sum\limits_{k=1, k\neq l}^{N_g} \sum\limits_{l\neq k}^{N_g} \sum\limits_{q=1}^{M} A^{**}_{kl(q)}(\tau) S_{kl,d,n,c}}, \tag{3}$$

where $S_{kl,d,n,c}$ allows for the selection of a subset of the network; for example, for dayside connections $S_{kl,d} = 1$ if both grid cell $k$ and $l$ lie on the dayside and $S_{kl,d} = 0$ otherwise. We also define parameters for the average connection likelihood between high-latitude ($\theta_h$) (MLAT > 66°), low-latitude ($\theta_l$) (MLAT < 66°), and cross-latitude ($\theta_c$) grid cells by selecting the relevant subsets of the network.

We also defined short-range connections as having geodesic separation <4000 km and long range as having a separation >4000 km; the $k$th then has normalized degree $n_{k,S}$ and $n_{k,L}$, respectively.

$$n_{k,S,L}(\tau) = \frac{\sum\limits_{l=1, l\neq k}^{N_g} \sum\limits_{q=1}^{M} A^*_{kl(q)}(\tau) S_{kl,S,L}}{\sum\limits_{l=1, l\neq k}^{N_g} \sum\limits_{q=1}^{M} A^{**}_{kl(q)}(\tau) S_{kl,S,L}}, \tag{4}$$

where $S_{kl,S,L}$ allows for the selection for short-range and long-range grid cell pairs. Network parameters $n_{k,S}$ and $n_{k,L}$ quantify how likely grid node $k$ is to be connected to spatial regions within 4000 km and beyond 4000 km. The normalized grid cell degree $n_k$ is formed by removing the $S_{kl}$ matrix from equation (4).

We obtain all parameters representing the aggregated networks as deviations from a baseline network $B_{kl}$. The baseline network captures the typical correlation between regions of MLT-MLAT that is seen when we average over all possible unconstrained quiet time IMF conditions. The baseline is obtained by averaging over two long time intervals on either side of the turning. These finite intervals are at 10,000–5000 min before, and 5000–10,000 min after the turning. The baseline networks are calculated from the averaged time-evolving network, that is, the networks derived after averaging over a given event set.

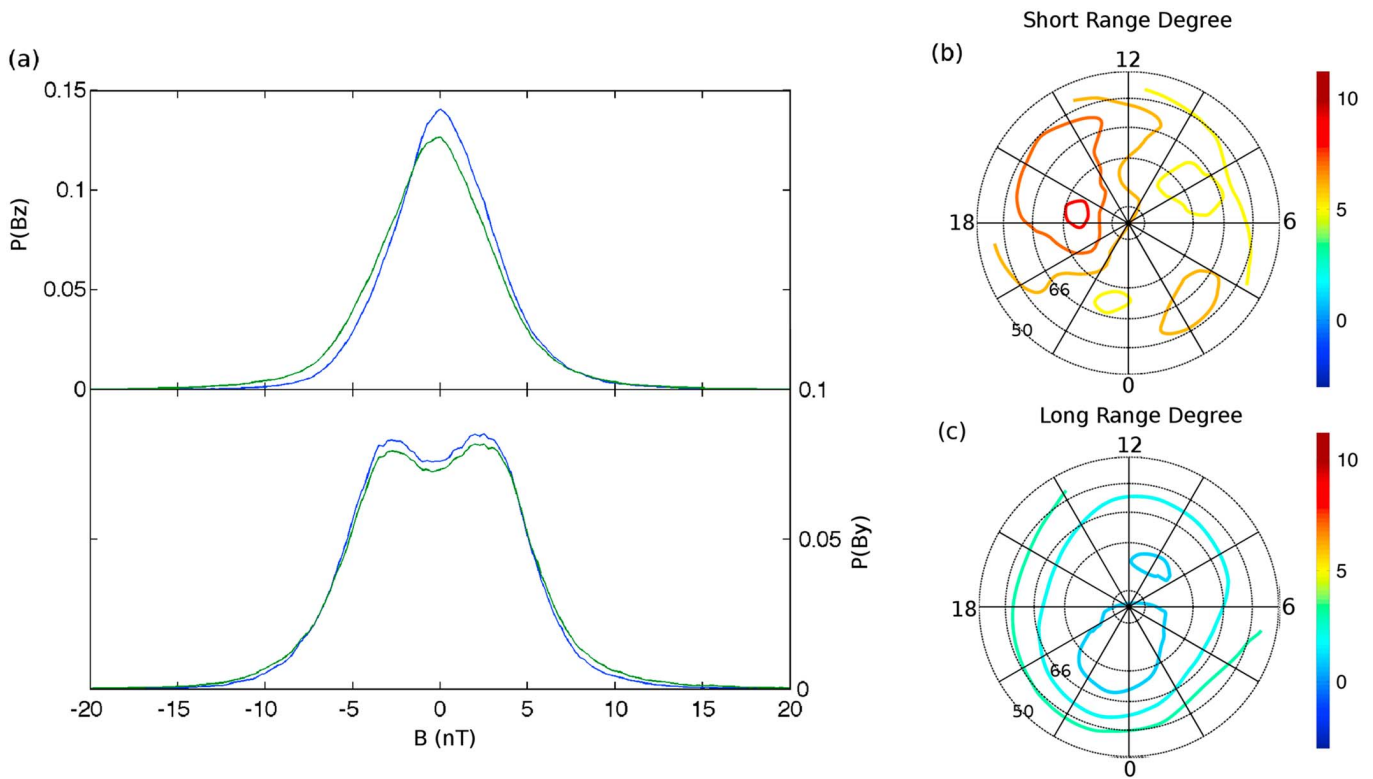

**Figure 1.** (a) The probability densities for IMF $B_z$ (top) and $B_y$ (bottom). The blue curves represent the probability densities during quiet time, and the green curves represent the probability density at all times. The contour plot shows an example of the baseline network showing (b) the short-range degree and (c) the long-range degree. The contour values represent the percent likelihood that a given region is connected to any other region in its network domain, that is, for short-range degree, a region's domain contains all the other regions <4000 km away. The redder the contour the greater the degree.

$$B_{kl} = \frac{\sum_{j=2500}^{5000} \sum_{q=1}^{M} A_{kl(q)}^{*}(j\Delta)}{\sum_{j=2500}^{5000} \sum_{q=1}^{M} A_{kl(q)}^{**}(j\Delta)}. \tag{5}$$

Where $\tau = j\Delta$ and $\Delta = 2$ min (as the time step is 2 min). A unique baseline network is formed for each of the aggregated north-south and south-north turning event sets.

In Figure 1a we show the probability densities for IMF $B_y$ and $B_z$, $P(B_y)$, and $P(B_z)$, respectively, during quiet time; that is, there are no substorms or storms and $|DST| < 30$. We can see that the negative tail of the $P(B_z)$ is reduced during quiet time conditions, with the positive tail largely unchanged. The $P(B_y)$ distribution shows that the probability of $|B_y| > 5$ nT is reduced during quiet time conditions. In Figures 1b and 1c we also plot an example of a baseline network, Figure 1b shows the short-range degree and Figure 1(c) the long-range degree. The short-range degree shows that there is more short-range correlation between regions at 12–20 MLT and between regions at 00–04 MLT. The location of the increased correlation coincide with the expected locations of a twin convection cell, albeit under strongly +$B_y$ conditions. However, for the baseline networks the average $B_y \sim 0$ and $B_z \sim 0$, indicating that there is some inherent asymmetry.

## 4. Correlation Network Response to IMF Turnings

Figure 2 shows aggregate network parameters plotted as a function of the time that has elapsed (delay) $\tau$ following the north-south and south-north turnings of the IMF propagated to the magnetopause. Each subfigure, from top to bottom, plots as follows: $\alpha$, the average connectivity over all turning events and all grid cells in the network; the average connectivity between the dayside regions $\phi_d$ (green), the nightside regions $\phi_n$ (blue), and connections between the dayside and nightside regions $\phi_c$ (red); the average connectivity between high-latitude regions $\theta_h$ (green) and low-latitude regions $\theta_l$ (blue), and between high- and low-latitude regions $\theta_c$ (red); and the bottom panel in each of the subfigures plots the IMF $B_y$ and $B_z$ components averaged over all events used to form each aggregate response. The network parameters are always plotted at the time of the leading edge of the correlation window, i.e., at $\tau = 0$ data from $\tau = -48$ to 0 min are used to calculate the correlation network. The key results are presented in sections 4.1 and 4.2. The results for the network responses to IMF turnings that occur during the half year centered on winter or summer solstice and the influence of magnitude of $B_z$ are shown in section 4.3 and 4.4, respectively.

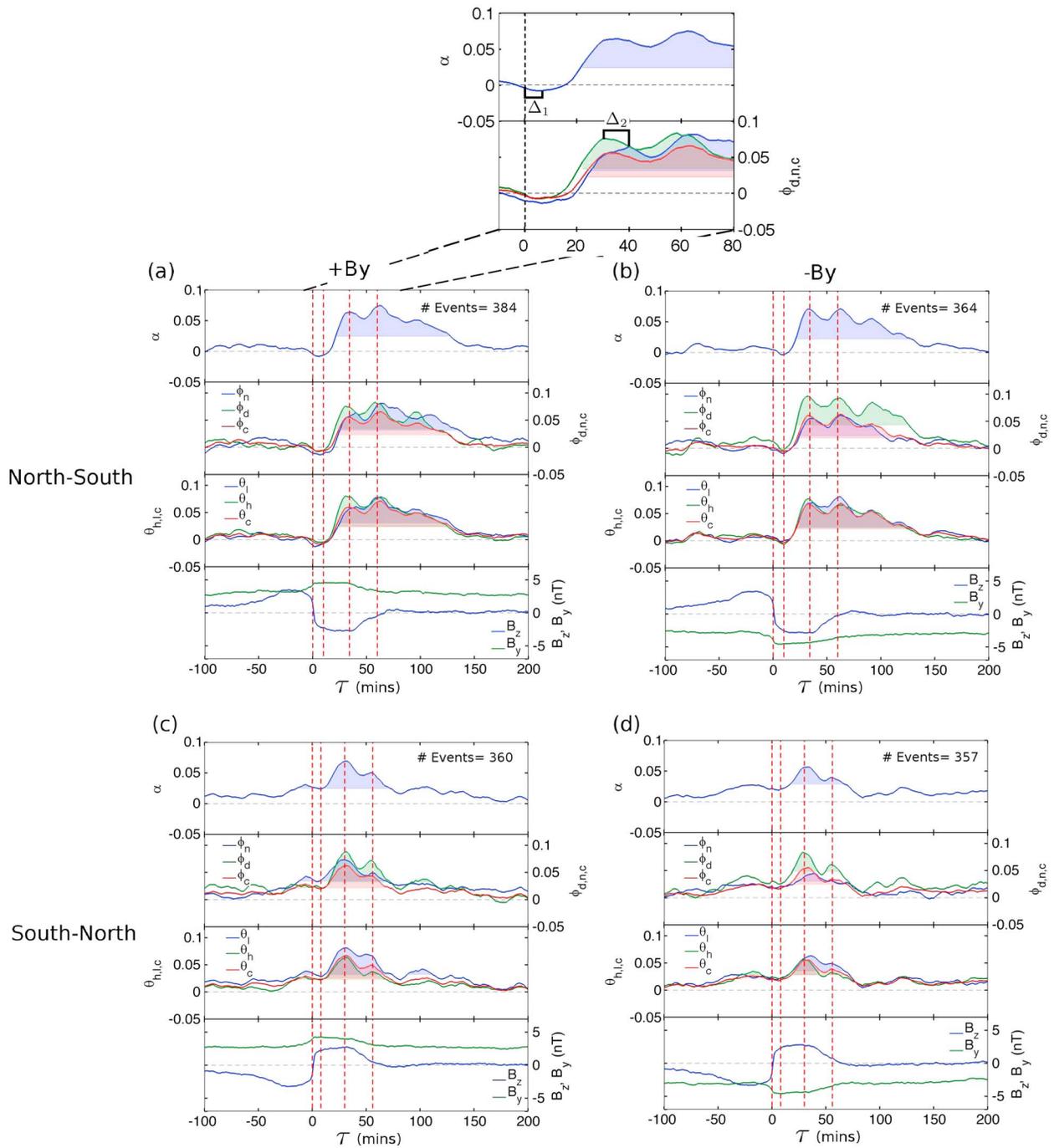

**Figure 2.** The aggregated network parameters for (a, b) north-south and (c, d) south-north IMF turnings for $+B_y$ (Figure 2a) and $-B_y$ (Figure 2c) conditions as a function of delay $\tau$ elapsed since the turnings. These results aggregate events where $B_y$ and $B_z$ do not change sign over a 40–80 min time interval post IMF turning. All parameters represent differences between the aggregate response for each subset of turning events and the baseline network. Each subfigure plots, from top to bottom, the following: $\alpha$, the average connectivity between all turning events and all grid cells in the network; the average connectivity between the dayside regions $\phi_d$ (green), the nightside regions $\phi_n$ (blue), and connections between the dayside and nightside regions $\phi_c$ (red); the average connectivity between high-latitude regions $\theta_h$ (green), low-latitude regions $\theta_l$ (blue), and between high- and low-latitude regions $\theta_c$ (red); the average IMF $B_z$ (blue) and $B_y$ (green) for the turning events. The shading under the curves indicates statistical significance; values above significance have a probability of <0.0001 by being generated by a random distribution. This random distribution was formed by considering a model for the distribution of the values of each of the parameters at $|\tau| > 5000$ min. Marked on all plots (vertical dashed red line) are several delays for which the network maps are displayed in Figure 3. The delays are $\tau = 0$, 10, 34, and 60 min for the north-south turnings (Figures 2a and 2b) and 0, 8, 30, and 56 for the south-north turning (Figures 2c and 2d). We also plot an expansion of a subset of the network parameters for the north-south turning $+B_y$ event set. Labeled on the plots are two time intervals $\Delta_1$ and $\Delta_2$. $\Delta_1$ is the time between the turning reaching the magnetopause and the initial network response, determined by the time it takes before $\alpha$ begins to increase continuously into a significant peak post turning. $\Delta_2$ is the time interval between the first peak response in $\phi_d$ and the first peak response in $\phi_n$.

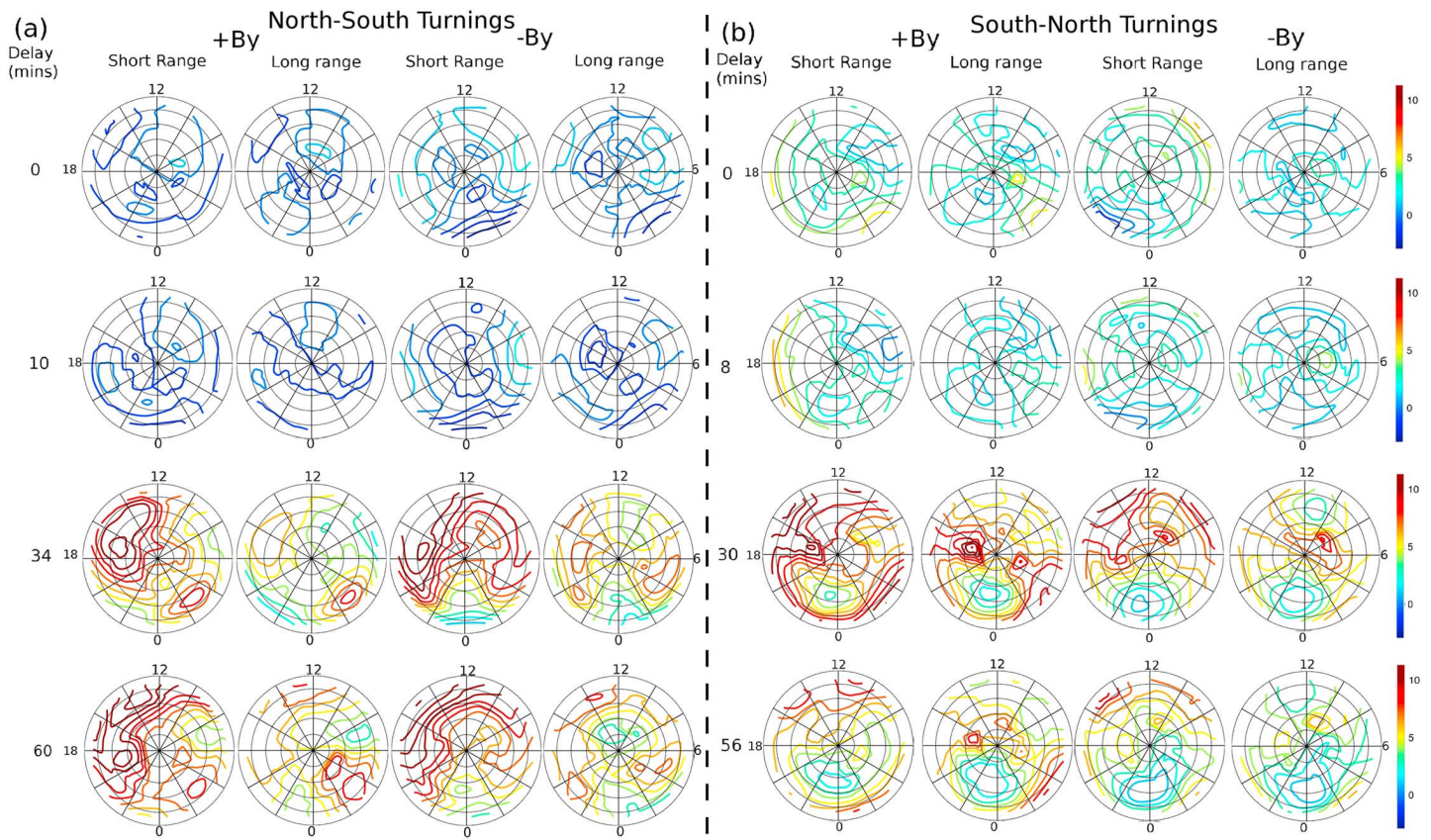

**Figure 3.** Snapshots of the correlation network maps at the delays $\tau$ following the turnings highlighted by the vertical dashed lines in Figure 2 for (a) north-south and (b) south-north turnings. Each subset of the panel is as follows. (left column) The short range degree which quantifies extent of connection between a particular region and all of its close neighbors (regions within 4000 km of each other). (right column) The long-range degree which quantifies the extent of connection between a given region and all other distant regions (regions greater than 4000 km away from each other). The contour values represent the percent likelihood, relative to baseline, that a given region is connected to any other region in its network domain. The redder the contour the greater the increase in degree. The black dotted concentric circles represent the MLAT contours. They are from outer to inner 50°, 58°, 66°, 74°, and 82° MLAT. While contours are plotted at latitudes greater than 82°, no data are taken in this region.

### 4.1. IMF $B_z$ North-South Turnings

Regardless of the duration of the southward IMF post turning, all event sets share a similar overall aggregate network response to the north-south turnings. At $\tau = 0$ min, Figures 2a and 2b show that the network response dips slightly below baseline, with reduced correlation primarily at low latitudes. Figure 3a provides a visual representation of the correlation network relative to baseline at several times that are indicated by the red vertical dashed lines in Figure 2. The contour values represent the percent likelihood, relative to baseline, that a given region is connected to any other region in its network domain, that is, for short-range degree, a region's domain contains all the other regions <4000 km away. At $\tau = 0$ following the IMF turning, the dark blue regions of the degree maps show where the correlation is reduced, which is predominantly at low latitudes for both the $+B_y$ and $-B_y$ north-south event sets, for long- and short-range correlation. Below baseline correlation may indicate either weak ambient current system and/or a fast changing system that varies over small spatial scales that are not resolved by the grid.

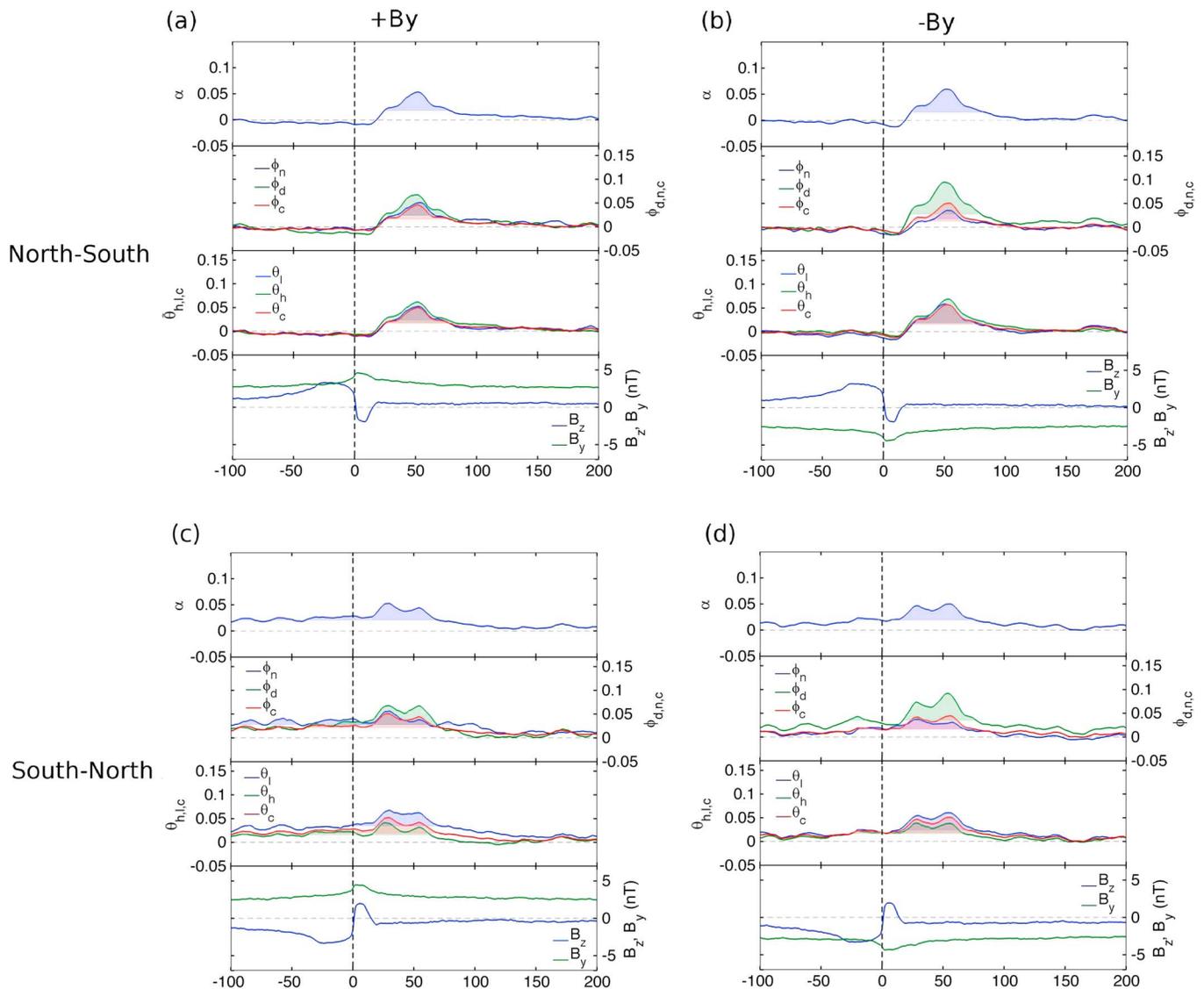

**Figure 4.** The aggregated network parameters for (a, b) north-south and (c, d) south-north IMF turnings for $+B_y$ (Figures 4a and 4c) and $-B_y$ (Figures 4b and 4d) conditions as a function of time delay since the turning for the 10–20 min interval range. The figure follows the same format as Figure 2. The time difference between the maximal response in $\phi_d$ and $\phi_n$ is 0–2 min for all 10–20 min interval event sets.

Figures 2a and 2b show that after $\tau = 10 \pm 2$ min, correlation in the network begins to increase, indicating that the magnetometer response to the turnings begins to fall within the leading edge of the correlation window. This suggests that 10 min is the average communication delay between the time at which the southward turning reaches the magnetopause and the onset of the spatiotemporally correlated magnetometer response. Correlation in $\phi_d$ that is, the extent of correlation between regions within the dayside, begins to increase ~2–8 min before that of $\phi_n$ that is, the extent of correlation between regions within the nightside.

Shorter and longer post turning intervals were considered to investigate the effect of the length of continually negative IMF $B_z$ on the aggregated network response. Figures 4 and 5 show the results for the 10–20 and 20–40 min interval ranges using the same methods as those for the 40–80 min interval range. They show shorter delays between increases in $\phi_d$ and $\phi_n$ of ~2 min. The response in $\phi_c$ that is the extent of correlation between dayside and nightside regions, is typically seen after $\phi_d$ but before $\phi_n$ by ~2 min.

For the longest interval range 80–150 min, Figure 6, there is an insufficient number of events to divide the north-south and south-north turnings into $+B_y$ and $-B_y$ event sets. Figure 6a shows similar behavior to the 40–80 min interval north-south turning event sets (Figure 2). The main differences between these are the increased length of time that $\phi_n$ remains above significance which is consistent with the longer post turning southward directed field, and the time difference in the response of $\phi_d$ and $\phi_n$ is 8 min, which is longer than the other event sets.

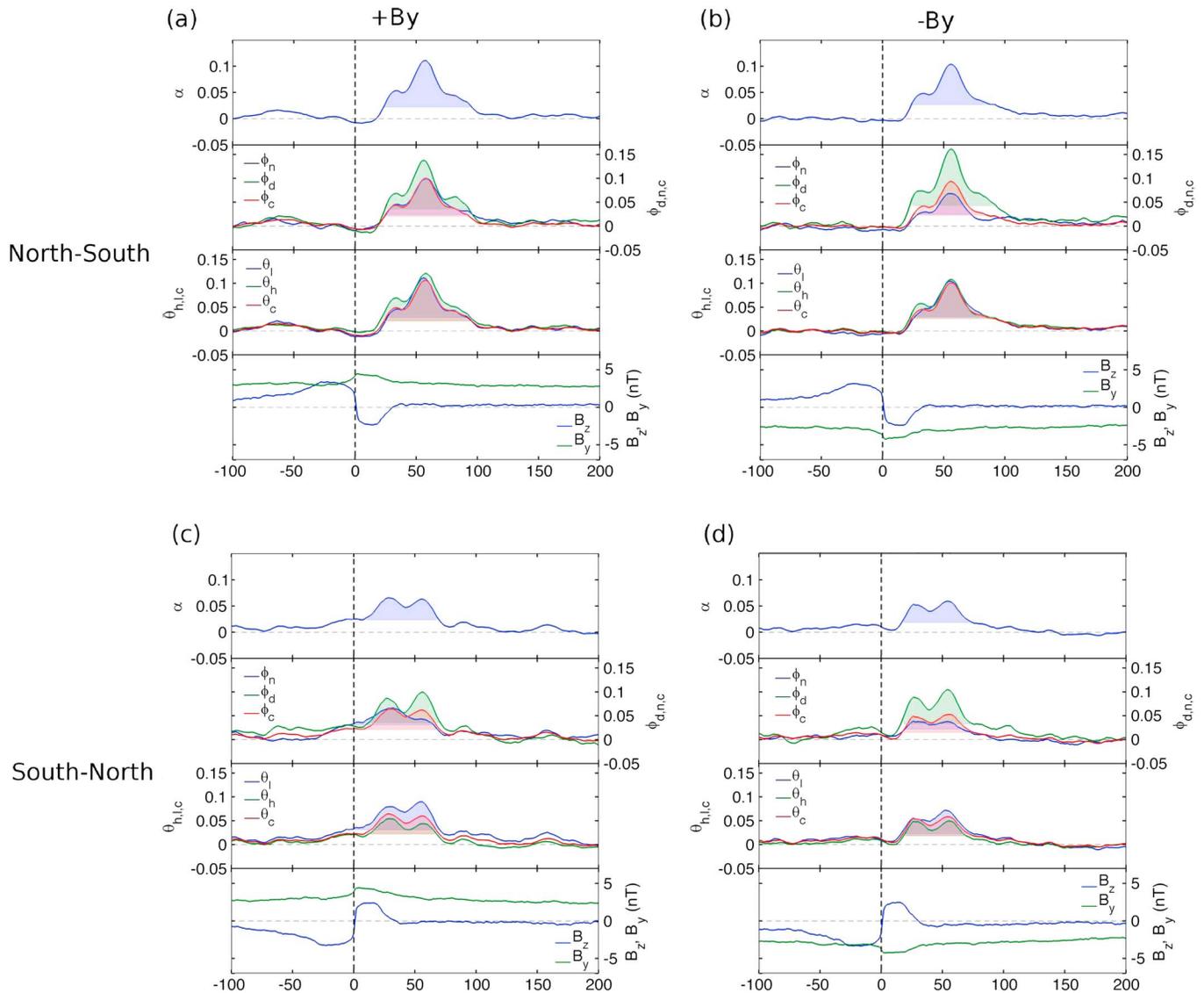

**Figure 5.** The aggregated network parameters for (a, b) north-south and (c, d) south-north IMF turnings for $+B_y$ (Figures 5a and 5c) and $-B_y$ (Figures 5b and 5d) conditions as a function of time delay since the turning for the 20–40 min interval range. The figure follows the same format as Figure 2. The delay between the maximal response in $\phi_d$ and $\phi_n$ is 0–2 min for all 20–40 min interval event sets.

For all turning sets there is a multiple peak structure in $\alpha$ with the first and second peaks occurring at $\tau = 34 \pm 2$ and $\tau = 60 \pm 2$ min. The correlation window is 48 min and multiple subpeaks may simply indicate time structure that is not resolved; see Appendix B for a simple model that clarifies this. Figure 3a shows the degree maps at $\tau = 34$ min for the $+B_y$ and $-B_y$ event sets. The duskside has larger increases in short-range correlation at $\tau = 34$ min for both the $+B_y$ and $-B_y$ event sets when compared to the increases on the dawnside. The long-range degree does not show the same asymmetry. Additionally, regions of increases in short and long-range degree in the dawn hemisphere are shifted more toward dayside and spread over a wider area, 03–10 MLT, for the $-B_y$ set compared to 01–05 MLT in the $+B_y$ set. The differences in the spatial locations of the responses for the $+B_y$ and $-B_y$ event sets may reflect the rotation of the current system pattern. Figure 3a shows that regions that experience large increases in long- and short-range correlation roughly coincide with the expected locations of negative and positive convection cells during southward IMF. The orientation of the dawn-dusk convection cells is known to be influenced by the sign of $B_y$, with rotation of the system clockwise for $+B_y$ and vice versa [*Walsh et al.*, 2014]. A similar rotation can be seen in our results, with the dawn region of increased long-range correlation being shifted toward the dayside for the $-B_y$ set compared to the $+B_y$ set. This suggests that the correlation network is highlighting aspects of the convection system response.

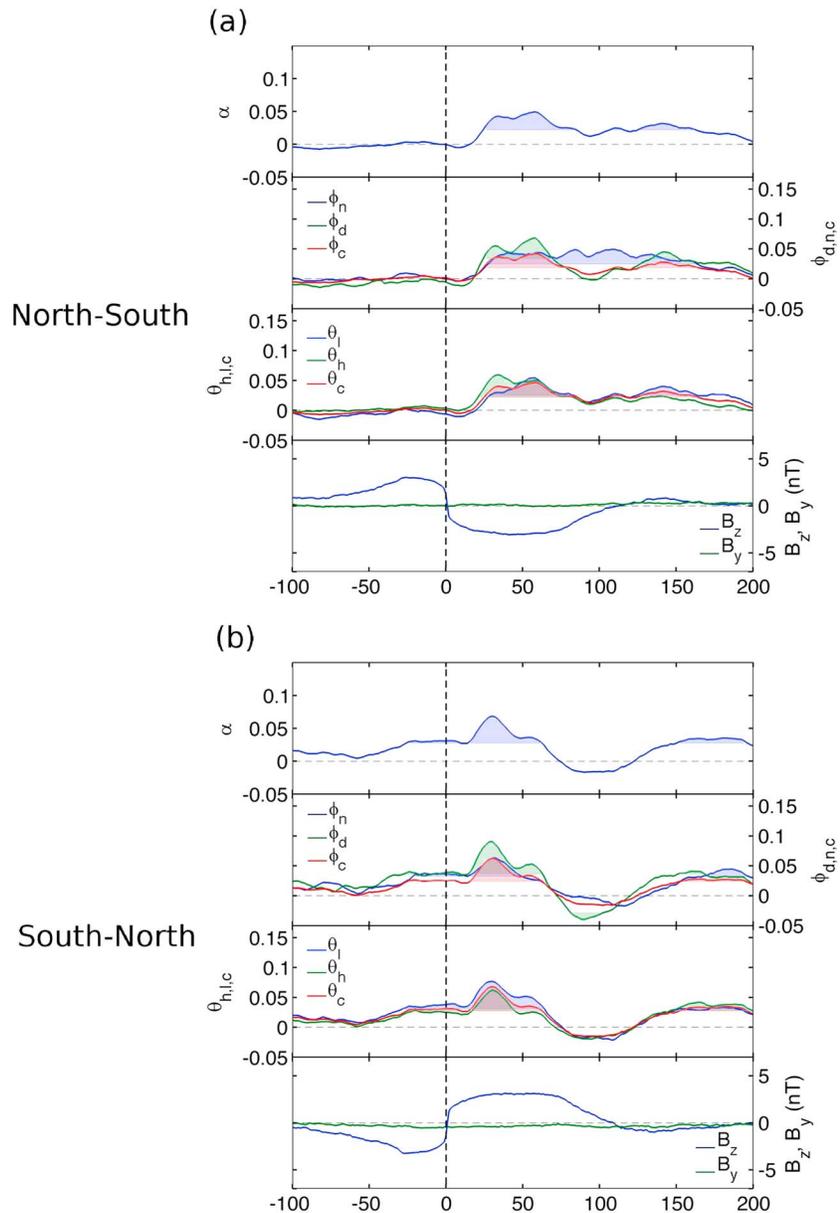

**Figure 6.** The aggregated network parameters for (a) north-south and (b) south-north IMF turnings for the 80–150 min interval range. The figure follows the same format as Figure 2. The delay between the maximal response in $\phi_d$ and $\phi_n$ is 2 min for the south-north event sets and 8 min for the north-south event sets.

At $\tau = 60$ min, for the $+B_y$ event set, Figure 3a shows that the regions of elevated short- and long-range correlation have expanded to lower latitudes as well as covering a larger area in MLT. There is no clear expansion of the regions of strong correlation to lower latitudes for the $-B_y$ event set in the dawnside. After $\tau \simeq 60$ min the IMF $B_z$ averages to zero for both event sets. Correlation in the network remains above the significance level until $\tau \simeq 124$ min for both event sets. Since the time is plotted from the leading edge of the window, this corresponds to correlation decreasing at $\tau \geq 76$ min, that is, at the trailing edge of the correlation window. Also in the $+B_y$ set, $\phi_d$ drops below significance before $\phi_n$ and the opposite is true for the $-B_y$ case. The difference in dayside correlation in the $+B_y$ and $-B_y$ event sets may be due to the rotation of the equivalent current system. The duskside typically exhibits larger increases in correlation. For the $+B_y$ set the increase in correlation on the duskside no longer spans the day and night hemispheres and is mostly in the dayside. The raised response in the network after the average $B_z$ has fallen to zero could indicate that the energy stored during the period southward IMF slowly dissipating over a timescale $\sim 60$ min. Our events only include quiet time nonsubstorm times; hence, our proposed dissipation timescale applies to these times only.

### 4.2. IMF $B_z$ South-North Turnings

Figures 2c and 2d show the average network response to the south-north turnings events for $+B_y$ and $-B_y$, respectively. As in the case of the north-south turnings, correlation begins to increase at $\tau \sim 8$ min, suggesting a delay of $\sim 8$ min between the time that the turning reaches the magnetosphere and the network response for the south-north turnings. This could suggest a magnetopause-ionosphere information transit time for both the north-south and south-north turnings that are similar to that found by Ridley et al. [1998, and references therein].

In contrast to the north-south turnings, from $\tau = 0$ to $\tau = 8$ min, correlation in the network is enhanced above baseline for both $+B_y$ and $-B_y$ event sets. Figure 3b provides snapshots of the network in an identical format to Figure 3a. The above baseline network response at the turning, $\tau = 0$, is mostly due to short-range correlation at low latitudes. The picture is largely the same at $\tau = 8$ min for both $+B_y$ and $-B_y$ event sets.

As we found for the north-south turnings, Figures 2c and 2d show the same multiple peak structure in $\alpha$, but this occurs sooner than in the north-south turnings, with peaks at $\tau = 30$ and 56 min in comparison to the $\tau = 34$ and $\tau = 60$ min for the north-south turnings. The short-range correlation at these times is strongest at low latitudes and long range at high latitudes. For the south-north turnings, responses in $\phi_n$ are already raised above significance before the turning in the $+B_y$ event set which is probably due to southward IMF, exciting the system before the turning. As such, we could not obtain an estimate of the delay between responses in $\phi_d$ and $\phi_n$ in this case and others post turning intervals. In the $-B_y$ event set there does seem to be a delay between the initial responses of $\phi_d$ and $\phi_n$ of $\sim 6$ min. Again, $\phi_c$ responds $\sim 2$ min after $\phi_d$.

For the post turning intervals of $40-80$ and $80-150$ min, shown in Figure 6, our results for both the north-south and south-north turnings, shown in Figure 2, show delays in response times of $2-8$ min in $\phi_d$, the correlation within the dayside region, and $\phi_n$, the correlation within the nightside region. The correlation between day and night regions ($\phi_c$) typically responds $\sim 2$ min after $\phi_d$ for the same event sets. However, the event sets with shorter post turning intervals shown in Figures 4 and 5 typically show shorter delays, $\sim 2$ min, between $\phi_d$ and $\phi_n$.

At $\tau = 30$ min, in Figure 3b, we can see that in the $+B_y$ event set the short-range correlation shows the strongest increase at low-latitude regions at $12-19$ MLT and $2-5$ MLT. Increases in long-range correlation are more local, with the strongest increases across high latitudes at $16-19$ MLT and $03-06$ MLT. In the $-B_y$ set, increases in short range correlation are largely confined to the dayside. The strongest increases in long range correlation occur at high latitude on the dawn hemisphere at $04-10$ MLT with smaller increases occurring at $14-18$ MLT. In both events sets the $22-02$ MLT region shows almost no significant response in long- and short-range correlation (blue regions). These blue regions again indicate either no response or unresolved fine spatial structure and/or fast timescales. As in the north-south turnings, increases in correlation appear to spatially coincide with regions where the two-cell convection pattern would be present given $+B_y$ or $-B_y$.

Figure 2 shows that the second peak in correlation at $\tau = 56$ min is dominated by increases in $\phi_d$, with only a slight increase across dayside-nightside and within the nightside correlation for the $+B_y$ case. In Figure 3b the response seen at $\tau = 30$ min has started to die away, with some remaining correlation at low latitudes and some small islands at high latitudes in long-range degree. The blue regions indicating no enhanced correlation has expanded. For the south-north $80-150$ min post turning interval set we found that, as shown in Figure 6b, there are novel features not seen at the other interval ranges. There is a significant decrease in correlation after $\tau = 56$ min, reaching a minimum at $\tau = 90$ for dayside correlation with a minimum for nightside correlation 26 min after. This dropout in correlation to below baseline is not seen in any of the shorter turning intervals, possibly because the IMF has not remained northward for a long enough period of time.

### 4.3. Summer and Winter Centered North-South and South-North Turnings

The north-south and south-north turning events are subdivided into sets that occur during the half year centered on winter or summer solstice. Figure 7 provides the results for the north-south and south-north turning events in the same format as Figure 2.

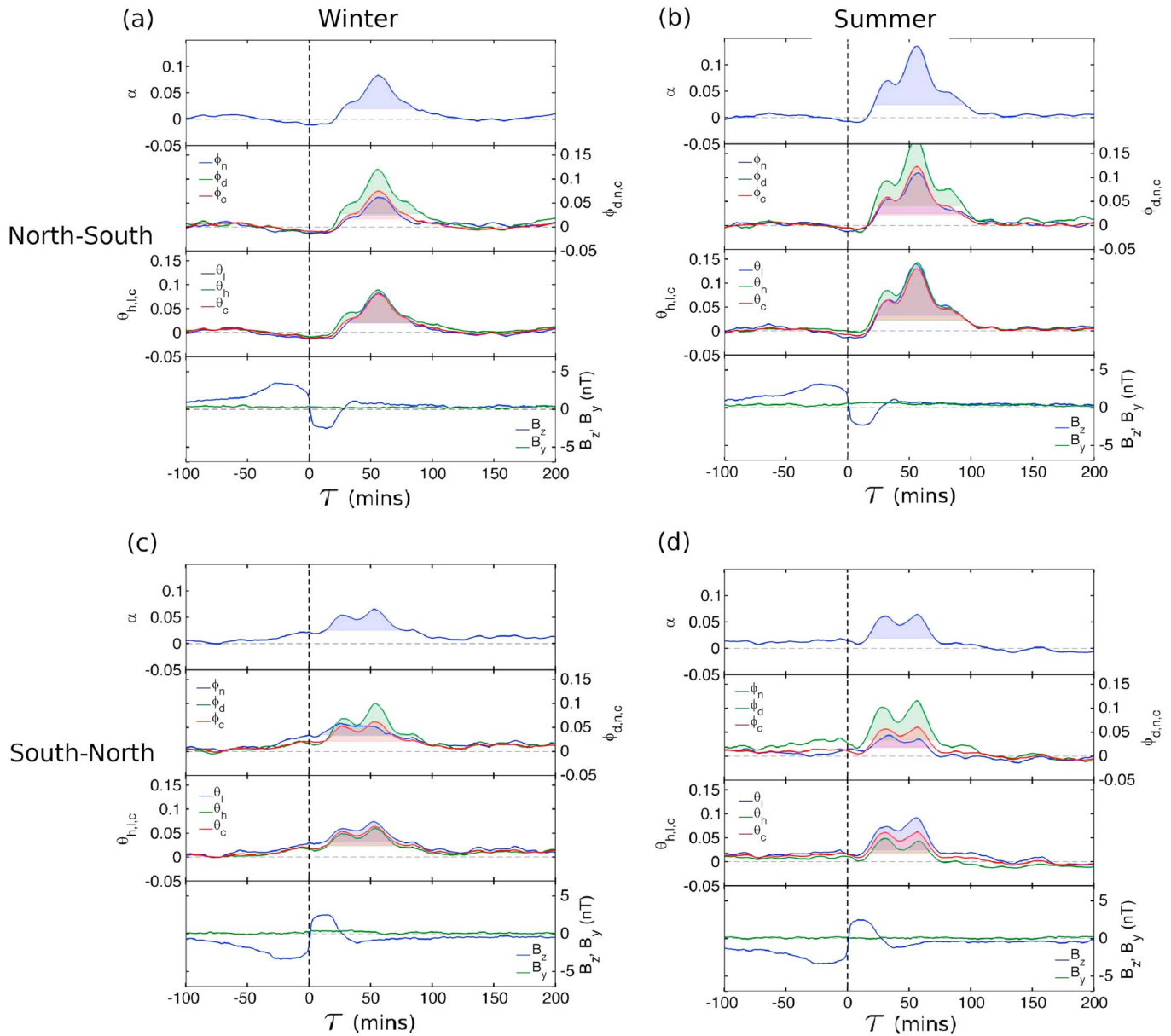

**Figure 7.** The aggregated network parameters for (a, b) north-south and (c, d) south-north IMF turnings for winter (Figures 7a and 7c) and summer (Figures 7b and 7d) conditions as a function of time delay since the turning for the 20–40 min interval range. All parameters represent differences between the aggregate response for each subset of turning events and the baseline network. The summer events show increased correlation overall, with the majority of the increases coming from increases dayside correlation.

We can see that there are two notable differences between the winter and summer events. The overall response to the turning ($\alpha$) is larger for the summer events, with the majority of the increased correlation arising from increases in dayside correlation ($\phi_d$). Additionally, the south-north turnings shows a stronger response in the low-latitude correlation ($\theta_l$) during summer events. Both of these observations suggest that the ionospheric conductivity due to sunlight, known to have an influence on the equivalent currents resulting from convection [*Laundal et al.*, 2016], also has an influence the correlation between regions of the system.

### 4.4. Strong and Weak IMF Turnings

The north-south and south-north turning events are subdivided into sets where the post turning IMF $|B_z|$ is continually <2 nT for the "weak" turnings events and >2 nT for the strong turning events. We did not explicitly control for $|B_z|$ before the turnings nor were $B_y$ controlled for post turning due to an insufficient number of events.

**Figure 8.** The aggregated network parameters for (a, b) north-south and (c, d) south-north IMF turnings for the 20–40 min post turning interval range. The north-south and south-north turning events are split into those where post turning $|B_z|$ is continually <2 nT post turning (Figures 8a and 8c) and >2 nT for the strong turning events (Figures 8b and 8d). The strong event sets show a significantly larger response to the turnings and have time differences between the maximums in $\phi_d$ and $\phi_n$ of 0–2 min compared to 4–8 min for the weak turning event sets.

We can see from Figure 8 that there is a significantly reduced response for both the weak north-south and south-north turnings events. Additionally, delays between responses in $\phi_d$ and $\phi_n$ are >4 min for the weak turnings and ≤2 min for the strong turnings. This may suggest that the speed of convective change is dependent on the magnitude of the IMF turnings. Alternatively, the results could be interpreted in the context of the two step process mentioned before [*Lu et al.*, 2002b]. If the correlated magnetometer response magnitude reflects both a magnetosonic wave launched at the magnetopause by the southward turning and the slower reconfiguration of the transient equivalent currents. The former may depend strongly on the magnitude of the turning whereas the latter does not.

If the magnitude of the response in the magnetometers due to a magnetosonic wave launched by the southward turning is dependent on the magnitude of the turning, and if the magnetometer response due to the slower reconfiguration of the transient equivalent currents is not as strongly dependent, this would lead to a magnetometer response dominated by the reconfiguration in the weak case and the wave in the strong turning case, leading to respective delays seen in correlation network response.

## 5. Discussion

Our results are relevant to several outstanding questions on the magnetospheric-ionospheric response to IMF turnings. We investigated the spatial and temporal timescales of the transient currents produced by the IMF $B_z$ turnings. There is currently support for two competing concepts for how the current systems in the ionosphere change following an IMF $B_z$ turning. The first is a fast initiation of the transient currents or ion flow associated with the turning being felt at all MLTs in the high-latitude ionosphere almost simultaneously (<2 min) [*Ridley et al.*, 1997, 1998; *Ruohoniemi and Greenwald*, 1998; *Yu and Ridley*, 2009]. The second is a gradual reconfiguration, where the response to the IMF turning is first seen in the dayside and then more gradually spreads to the nightside [*Lockwood et al.*, 1986; *Todd et al.*, 1988; *Khan and Cowley*, 1999; *Fiori et al.*, 2012; *Anderson et al.*, 2014]. Many of theses studies report a delay of ~10 min between a response at the dayside cusp region and a response at midnight. However, most recently, *Anderson et al.* [2014] found a much longer delay of 30–40 min between observations of FAC currents on the dayside and the nightside. Care must be taken in any comparison between the results as different sets of observations are used; *Anderson et al.* [2014] used AMPERE Iridium satellite data to measure the FAC in the dayside and nightside, whereas, e.g., *Fiori et al.* [2012] and others [*Lockwood et al.*, 1986; *Todd et al.*, 1988; *Khan and Cowley*, 1999] use a mix of local ground-based radar and magnetometer measurements. Importantly, the nightside FAC current response in *Anderson et al.* [2014] is likely associated with the onset of a substorm, where the nightside currents first increase around midnight and then expand toward the dayside. Conversely the "slow reconfiguration" studies [*Lockwood et al.*, 1986; *Todd et al.*, 1988; *Khan and Cowley*, 1999; *Fiori et al.*, 2012; *Anderson et al.*, 2014] find evidence of a trackable expansion of reconfiguration signatures from the dayside cusp region toward the nightside region. In addition, *Anderson et al.* [2014] find a delay between the southward turning and an increase in dayside currents of 20 min, significantly longer than ~8 min delay found in the other studies cited.

Here we have found that the average delay between a turning reaching the magnetopause and any detectable network response was ~8–10 min, this is in good agreement with previous studies [see *Ridley et al.*, 1998; *Fiori et al.*, 2012, and references therein]. This estimate is at most as precise as the technique used to propagate solar wind values to the magnetopause [*Weimer et al.*, 2003]. Uncertainties associated with the constrained minimum variance technique of *Weimer et al.* [2003] are dependent upon the level of variance in the data (during steady periods the errors are larger). *Mailyan et al.* [2008] investigated the uncertainties in the propagation time for Weimer's technique for solar wind discontinuities. They found that 65% of the propagation delay estimates fell with ±5 min of the true propagation delay. Importantly, the errors are symmetrically distributed about zero, so that the errors would symmetrically smear any peaks in network connectivity from the aggregated networks.

We also found that the delays between a network response in the dayside and a response in the nightside differ between sets of events selected on the duration that IMF $B_z$ stays southward (or northward) following the turning; the "post turning" interval. We found delays of ~0–4 min between the dayside and nightside response for the 10–20 and 20–40 min post turnings interval ranges and ~2–8 min delay for the 40–80 and 80–150 min post turnings interval ranges. However, the peak total network response, $\alpha$, is the same for all post turning interval ranges. Since we only obtain a new network every 2 min and each network is derived from a 48 min window, these observation are at the edge of the resolution of the technique. The delays of ~0–4 min are more consistent with a rapid response of the ionosphere to the IMF turnings where the ~2–8 min delays for the longer post turning intervals are more consistent with a slower reconfiguration that begins in the dayside. As such, the correlation networks may be observing two distinct and related contemporaneous processes for the longer post turning interval ranges. *Lu et al.* [2002b] proposed that there are two main processes that occur after an IMF turning: a magnetoacoustic wave launched by the turning and the transient currents produced by the transfer of flux resulting from reconnection. It is suggested that this reconfiguration lasts ~15–25 min in terms of the large-scale topological changes, with the cross-polar cap potential continuing to intensify for another ~20 min [*Fiori et al.*, 2012]. In the context of our work we propose that the magnetoacoustic wave would account for both long range instantaneous correlation between day and night regions, that is, the response in $\phi_c$ and the shorter delays in the 10–20 and 20–40 min post turning interval events. Alternatively in the 40–80 and 80–150 min post turning interval event sets, a magnetometer response (given that the magnetometer is in the right position at the right time) will contain contributions from both the initial magnetoacoustic wave and the reconfiguration currents. To determine the delays between day and night responses, we used the time difference between the peak network response seen on the dayside and then the nightside. The time position of the peak network response depends on the relative contributions of the wave and the reconfiguration response to the network response. In the 40–80/80–150 event sets the correlation window will contain a bigger contribution from the reconfiguration currents, hence, the position of the peak will be shifted to a later time.

How we identified the north-south and south-north turnings differs from some other studies. Only sharp IMF turnings are used in *Ridley et al.* [1997, 1998], *Ruohoniemi and Greenwald* [1998], *Yu and Ridley* [2009], *Lockwood et al.* [1986], *Todd et al.* [1988], *Khan and Cowley* [1999], and *Fiori et al.* [2012]. There are no such restrictions on the individual events used here. How this may affect the resulting transient ionospheric current response is not immediately clear. If there is, as proposed in *Ridley et al.* [1998] and *Lu et al.* [2002b], an impulsive magnetoacoustic wave that is launched when the turning reaches the magnetopause, then the magnitude of this wave will depend on the sharpness of the turnings. In section 4.4 we indirectly explored this by selecting for "strong" and "weak" turning events. Our requirement that the $|B_z| > 2$ nT post turning ensures that the transition is sharper than when such requirements are not imposed. This could explain why the delay between dayside and nightside responses is ~0–2 min for the "strong" event sets and >4 min for the "weak" event sets.

We have presented a novel methodology for investigating the coupled solar wind-magnetospheric-ionospheric system with the potential to leverage new information from historic measurements. The technique is generic and in future work other data sets such as AMPERE [*Anderson et al.*, 2002; *Clausen et al.*, 2012] could in principle be incorporated. By selecting for times in which we know that a particular physical process is occurring (such as an IMF turning), we can obtain the network response corresponding to those physical processes. A one-to-one mapping between an observed dynamical network response and a specific physical interpretation could be performed by using physical models to generate a predicted spatiotemporal magnetic perturbation pattern. This pattern could then be quantified in terms of its time-dependent network properties which could then be directly and quantitatively compared to the data-driven network response.

## 6. Conclusions

We used a dynamical network of ground station magnetometers to quantify their collective, spatially correlated quiet time large-scale time-dependent response to north-south and south-north IMF turnings. This in turn reflects the dynamics of large-scale ionospheric currents and corresponding convection. The spatiotemporal correlation between the full set of SuperMAG ground station magnetometers between 50 and 82° MLAT in the Northern Hemisphere was used to construct the network. This method directly relies on the amplitude and phase information of all magnetometer time series to characterize the system. We constructed time-varying networks via canonical correlation between the vector time series of pairs of magnetometer stations at zero lag for the years 1998–2004. Network information was mapped onto a stationary regular grid in MLT and MLAT. We identified north-south and south-north turning events during quiet time conditions (times in which no storms or substorms are occurring) for which $B_y$ is either continually positive or negative after the turnings using ACE IMF data that had been propagated to the magnetopause. Aggregated network responses were formed for similar events as a function of delay ($\tau$) since the north-south and south-north turnings. We find the following:

1. Increases in spatiotemporal correlation in the network following a south-north and north-south turning. The detailed spatial pattern of this response depends on IMF $B_y$ and whether the turning is north-south or south-north.

2. There is a delay of ~8–10 min between the turning reaching the magnetopause and the response seen in the correlation network.

3. The strongest increases in short-range (geodesic separation between stations <4000 km) correlation is almost always in the afternoon-dusk region. Long-range (geodesic separation >4000 km) correlation does not show this bias.

4. The spatial pattern of long-range correlation is reminiscent of a two-cell convection pattern.

5. Correlation is absent in the south-north turnings at midnight suggesting either weak current systems and/or a unresolved highly structured, fast-changing system.

6. A 2–8 min delay between responses within the dayside ($\phi_d$, $6 < \text{MLT} < 18$) and responses within the nightside ($\phi_n$, $6 > MLT$ or $MLT > 18$). In addition, there is a significant response in correlation between dayside and nightside regions, $\phi_c$, that occurs shortly after $\phi_d$ and before $\phi_n$.

This work illustrates that dynamic correlation networks can characterize the spatiotemporal ionospheric response seen in the full set of ground-based magnetometer stations. We find the time between the turnings reaching the magnetopause and a network response to be ~8–10 min. If this time is comparable to the magnetopause-ionosphere information transit time, it is consistent with *Ridley et al.* [1998, and references therein]. In addition, we find tentative evidence for a two-step process in the transient equivalent current response to an IMF $B_z$ turning, that is, a fast initiation of the onset of change between day and night regions (evident from the smaller delay in response between $\phi_d$ and $\phi_n$ for shorter post turning intervals) followed by a more gradual reconfiguration occurring sooner on the dayside than the nightside (evident from the longer delays for the longer post turning intervals in which we expect significant convective change). In addition, we found, in section 4.4, that the strong turning events show delays of 0–2 min and the weak turnings, 4–8 min. Also, by comparing events occurring during summer and winter, see section 4.3, we find that ionospheric conductivity due to sunlight, known to have an influence on ionospheric equivalent currents [*Laundal et al.*, 2016], influences the strength of the correlation response in the dayside. Overall our results show that transitions in the equivalent current system occur coherently with significant long-range correlation between the expected locations of the convective cells. Our method could be used to perform detailed comparisons between the extensive sets of observation and dynamical models of ionospheric current systems, to identify the exact physical causes of correlation between regions.

## Appendix A: The Network Mapping Scheme

Here we describe the scheme for remapping the station correlation networks $T_{ij}$ onto a fixed regular grid in MLT-MLAT. The scheme is illustrated in Figure A1.

1. Each station, $i$, is assigned an area of influence (the blue bordered box in Figure A1). The area is same size, in degrees, as the grid cells and is centered on the station's location. The sizes of the grid cells (and thereby the station area of influence) in longitude and latitude are $\Delta\phi = 2$ h and $\Delta\theta = 8°$, respectively. The station network properties are then interpolated to the grid cells that overlap with the station's area of influence.

2. The weights for each cell node resulting from their neighboring stations are based on the percentage area overlap with each station. They are calculated as follows:

$$w_{ki} = (1 - \frac{\delta\phi_{ki}}{\Delta\phi})(1 - \frac{\delta\theta_{ki}}{\Delta\theta})\Theta(\Delta\phi - \delta\phi_{ki})\Theta(\Delta\theta - \delta\theta_{ki}), \tag{A1}$$

where $w_{ki}$ represents the weight for grid node $k$ resulting from station $i$, $\delta\phi_{ki}$ and $\delta\theta_{ki}$ the longitudinal and latitudinal separation between $i$ and $k$, respectively. $\Theta$ is the heaviside step function. This ensures that only the four closest nodes to station $i$ are included. By construction $\sum_{k}^{N} w_{ki} = 1$ except at the edges of the grid domain, i.e., if $\theta_i < 50 + \Delta\theta$, where $\theta_i$ is the latitudinal position of station $i$.

3. We then form two matrices $A_{kl}^{**}$ and $A_{kl}^{*}$. $A_{kl}^{**}$ is the maximum connectivity that could occur between grid cell pair $k$ and $l$, i.e., as if $T_{ij} = 1$ for all $ij$. $A_{kl}^{*}$ is the level of correlation between two grid cell pairs $k$ and $l$ resulting from correlation between stations $k$ and $l$. They are constructed as follows:

$$A_{kl}^{*} = \sum_{i=1,i\neq j}^{N_s}\sum_{j=1,j\neq i}^{N_s} w_{ki}w_{lj}T_{ij} \tag{A2}$$

and

$$A_{kl}^{**} = \sum_{i=1,i\neq j}^{N_s}\sum_{j=1,j\neq i}^{N_s} w_{ki}w_{lj}, \tag{A3}$$

where $N_s$ is the number of stations. Both $A_{kl}^{*}(t)$ and $A_{kl}^{**}(t)$ are a function of time. Figure A1b shows an example for one grid node pair $kl$ and three contributing stations. The station pair connection $T_{ij} = 1$ is shown with a blue dashed line and the calculated weights for grid node cell pair connection $A_{kl}^{*}$ and $A_{kl}^{**}$ are shown in black.

The mapping generates matrices $A_{kl}^{*}(t)$ and $A_{kl}^{**}(t)$: $A_{kl}^{**}$ is a normalization matrix and is the connectivity of grid cell pair $kl$ that would be obtained when all station pairs in the intermediate network are connected. $A_{kl}^{*}$ is the connectivity of grid cells $kl$ given by the time-dependent $T_{ij}$ network. Both $A_{kl}^{*}$ and $A_{kl}^{**}$ can take values $\geq 0$.

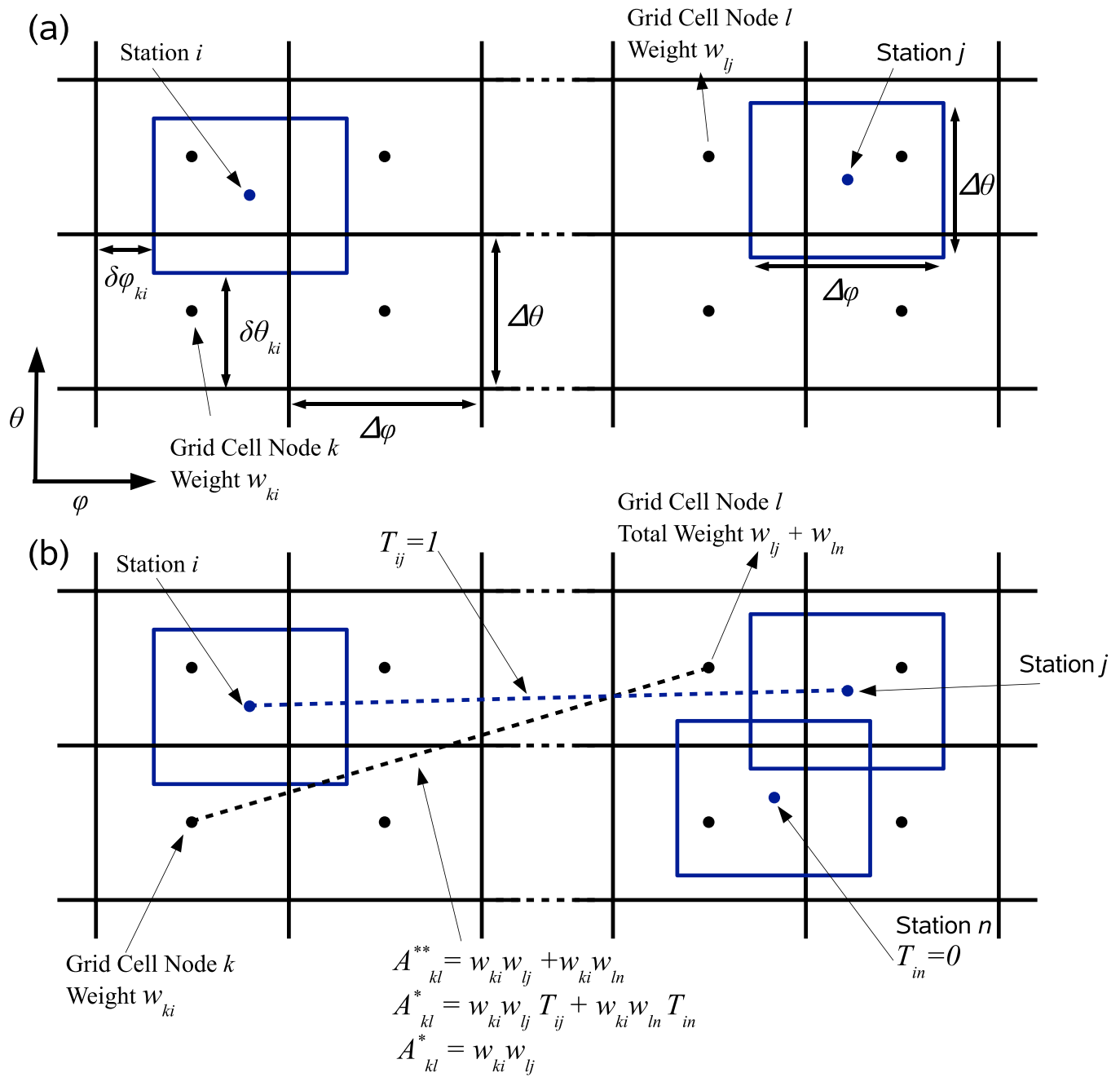

**Figure A1.** (a) A regular grid where the sizes of the grid cells are in latitude and longitude, $\Delta\theta$ and $\Delta\phi$, respectively, the station areas of influence and the separation between station $i$ and $k$ in MLT and MLAT, $\delta\phi$ and $\delta\theta$, are defined. (b) An example for calculating $A_{kl}^{**}$ and $A_{kl}^{*}$ given that there are only three neighboring stations $i$, $j$, and $n$. In this example stations $i$ and $j$ are connected and stations $i$ and $n$ are not connected, as such $A_{kl}^{*}$ receives no contributions from station pair $in$.

1. Each station, $i$, is assigned an area of influence (the blue bordered box in Figure A1). The area is same size, in degrees, as the grid cells and is centered on the station's location. The sizes of the grid cells (and thereby the station area of influence) in longitude and latitude are $\Delta\phi = 2$ h and $\Delta\theta = 8°$, respectively. The station network properties are then interpolated to the grid cells that overlap with the station's area of influence.

2. The weights for each cell node resulting from their neighboring stations are based on the percentage area overlap with each station. They are calculated as follows:

$$w_{ki} = \left(1 - \frac{\delta\phi_{ki}}{\Delta\phi}\right)\left(1 - \frac{\delta\theta_{ki}}{\Delta\theta}\right)\Theta(\Delta\phi - \delta\phi_{ki})\Theta(\Delta\theta - \delta\theta_{ki}), \quad (A4)$$

where $w_{ki}$ represents the weight for grid node $k$ resulting from station $i$, $\delta\phi_{ki}$, and $\delta\theta_{ki}$, the longitudinal and latitudinal separation between $i$ and $k$, respectively. $\Theta$ is the heaviside step function. This ensures that only the four closest nodes to station $i$ are included. By construction $\sum_{k}^{N} w_{ki} = 1$ except at the edges of the grid domain, i.e., if $\theta_i < 50 + \Delta\theta$, where $\theta_i$ is the latitudinal position of station $i$.

3. We then form two matrices $A_{kl}^{**}$ and $A_{kl}^{*}$. $A_{kl}^{**}$ is the maximum connectivity that could occur between grid cell pair $k$ and $l$, as if $T_{ij} = 1$ for all $ij$. $A_{kl}^{*}$ is the level of correlation between two grid cell pairs $k$ and $l$ resulting from correlation between stations $k$ and $l$. They are constructed as follows:

$$A_{kl}^{*} = \sum_{i=1, i\neq j}^{N_s} \sum_{j=1, j\neq i}^{N_s} w_{ki} w_{lj} T_{ij} \tag{A5}$$

and

$$A_{kl}^{**} = \sum_{i=1, i\neq j}^{N_s} \sum_{j=1, j\neq i}^{N_s} w_{ki} w_{lj}, \tag{A6}$$

where $N_s$ is the number of stations. Both $A_{kl}^{*}(t)$ and $A_{kl}^{*}(t)$ are a function of time. Figure A1b shows an example for one grid node pair $kl$ and three contributing stations. The station pair connection $T_{ij} = 1$ is shown with a blue dashed line and the calculated weights for grid node cell pair connection $A_{kl}^{*}$ and $A_{kl}^{**}$ are shown in black.

## Appendix B: Correlation Model for IMF Turnings

Here we show that the dual peak response seen in Figure 2 may simply arise from the finite length cross-correlation window undersampling the fine structure in the time series. To model the correlated magnetometer response, we use a signal $S_i(t) = \tanh(t) + R_i(t)$, where $R(t)$ is a randomized $1/f^{\alpha}$ noise (correlated noise) and $i$ references the test "stations," each of which has a unique noise signal $R_i(t)$. The tanh function is used to model an IMF turning. Ground station magnetometers typically have an $\alpha$ between 1 and 3 [*Jackel et al.*, 2001]; here we chose $\alpha = 2$ for all stations. The value of $\alpha$ will have an effect on the baseline correlation (the typical average correlation coefficient away from the turning) and increasing $\alpha$ will raise the baseline. We use cross correlation in lieu of canonical correlation used to form the real networks since we only have one component in this simplified model. We use an average power for the $R_i(t)$ noise component that is 10 times that of the tanh component signal. Forty-eight test signals $S_i(t)$ are constructed and the cross correlation between determined in an identical way to the methods found in section 2.

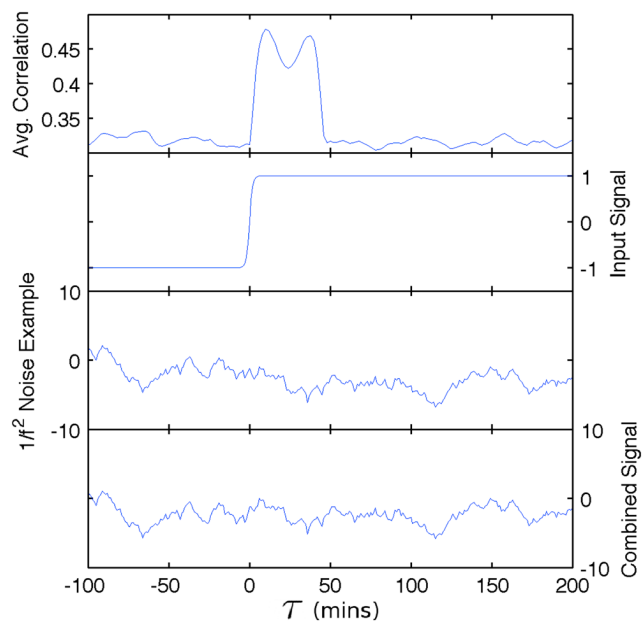

**Figure B1.** The figure from top to bottom shows the average cross correlation between all test signals as a function of delay since the turning, the tanh($t$) signal (exactly the same for all test signals), an example $R_i(t)$, and an example of the noise and the signal combined, $S_i(t)$.

Despite the test signals being noise dominated, Figure B1 shows that the model quantitatively recreates the same dual peak feature as found in Figure 2. Importantly, the separation between the peaks in the model is the same width as the separation between peaks in Figure 2. We conclude that the dual-peak response seen in Figure 2 may simply be a result of the windowing.

**Acknowledgments**
This work was supported by the UK Science and Technology Facilities Council, STFC project ST/K502418/1. The SuperMAG baselined ground magnetometer station data as well as the ACE solar wind data were obtained freely from http://supermag.jhuapl.edu/